# BINDING ENERGY OF DONOR STATES IN A QUANTUM DOT WITH PARABOLIC CONFINEMENT

S. Baskoutas[1], A. F. Terzis[2,*], E. Voutsinas[2],

[1] Department of Materials Science,
[2] Department of Physics,
University of Patras, Patras, GR-26500, Greece

**Abstract**

The donor binding energies associated with the ground state and a few excited states, are computed as a function of the dot size and the impurity position within two and three dimensional GaAs quantum dots. The calculation has been done using the Potential Morphing Method - a recently developed numerical method for the solution of time independent Schrödinger equation. The agreement with both perturbation and variational methods is very good, as regards the dependence of the binding energies on the dot size as well as on the location of the impurity within the quantum dot. Finally, we have shown that this method works well in all confinement limits - from weak to strong confinement.

*PACS*: 73.23. Ps; 73.20. Dx, 71.55. Eq

*Keywords*: Quantum dot; Impurity level; Semiconductor

---

Corresponding author: A.F. Terzis, email: terzis@physics.upatras.gr,

Tel.: +30 61099 7618, Fax: +30 61099 7618.

## 1. Introduction

It is well known that one of the main motivations behind the widespread interest in the physics of semiconductor heterostructures lies in the ability of producing quantum confined systems where carriers are restricted to move in two, one or zero dimensions (quantum wells, quantum wires and quantum dots (QDs), respectively). Due to the fact that impurities in semiconductors influence both transport and optical properties, topics like confined donors or acceptors in QDs have been extensively investigated [1-6].

In most theoretical investigations the confinement potential of QDs is assumed to possess square (either finite or infinite) or parabolic shape [7-9]. Many workers concentrated on the variational [2,7] or perturbation method [4,8-10], in order to calculate the donor binding energies for the ground and few adjacent states in QDs.

In the present work we will attempt to estimate also the binding energies for the ground and few adjacent states in two or three dimensional QDs with isotropic parabolic potential, using a recently developed numerical method for the solution of time independent Schrödinger equation, which is called potential morphing method (PMM) [11,12]. Comparison with the corresponding results of perturbation method [8] will point out the potentiality of the method, which has also the advantage (in comparison with the perturbation method) that it works well in all confinement limits (strong, medium, weak).

## 2. Theory

In the effective mass approximation, the Hamiltonian of a single hydrogenic impurity in a spherical QD (spherical symmetry) with parabolic confinement (second term) can be expressed as:

$$H = \frac{p^2}{2m^*} + \frac{1}{2}m^*\omega^2 r^2 - \frac{e^2}{\varepsilon|\vec{r} - \vec{r_i}|}, \qquad (1)$$

where $m^*$ is the electron effective mass, $e$ the electronic charge and $\varepsilon$ is the dielectric constant of the dot material. $\vec{r}$ is the position vector of the electron and $\vec{r_i}$ is the position vector of the fixed hydrogenic impurity.

Hence the time–independent Schrödinger equation is written as:

$$\left[-\frac{\hbar^2}{2m^*}\vec{\nabla}^2 + \frac{1}{2}m^*\omega^2 r^2 - \frac{e^2}{\varepsilon|\vec{r}-\vec{r}_i|}\right]\Psi(\vec{r}) = E\Psi(\vec{r}) \qquad (2)$$

Due to the fact that it is very convenient to work in dimensionless units we will write the above equation into the following form (see Appendix)

$$\left[-\vec{\nabla}_{(\vec{r}/R)} + \left(\frac{|\vec{r}|}{R}\right)^2 - \frac{2\left(\frac{R_y}{E_0}\right)^{1/2}}{\left|\frac{\vec{r}-\vec{r}_i}{R}\right|}\right]\widetilde{\Psi}\left(\frac{\vec{r}}{R}\right) = \widetilde{E}\widetilde{\Psi}\left(\frac{\vec{r}}{R}\right) \qquad (3)$$

Where $\widetilde{\Psi}\left(\frac{\vec{r}}{R}\right) = R^{d/2}\Psi(\vec{r})$, $d$ the dimensionality of the system under investigation and $\widetilde{E} = \frac{E}{E_0}$ is the energy in dimensionless units.

Usually for cases in which the potential is singular at the origin, we regularize it by the replacement $V(r) \rightarrow V(r + d_{cutoff})$ and later take $d_{cutoff} \rightarrow 0^+$. Therefore applying this procedure to the Coulomb potential we need to introduce a cut off distance $d_{cutoff}^2$ and the electrostatic interaction takes the following form [13,14]

$$\frac{e^2}{\varepsilon|\vec{r}-\vec{r}_i|} \rightarrow \left(\frac{e^2}{\varepsilon\sqrt{(x-x_i)^2 + (y-y_i)^2 + (z-z_i)^2 + d_{cutoff}^2}}\right)_{d_{cutoff}\rightarrow 0} \qquad (4)$$

We have defined this cut-off distance either through the mean square displacement ($d_{cutoff}^2 \equiv <|\vec{r}-\vec{r}_i|^2> - <|\vec{r}-\vec{r}_i|>^2$) or with a fixed value. The fixed $d_{cutoff}^2$ value found to be $10^{-5}\ R^2$. We specify this distance as the distance below which the mean value of the Coulomb potential does not changes. Obviously, for very low $d_{cutoff}^2$

values the mean value of the potential gives unphysical results due to computational round off errors.

Now in order to find eigenvalues and eigenfunctions of eq. (3) with the PMM [11,12] we need a reference system with well known eigenfunctions and eigenvalues, as for example the simple harmonic oscillator ($U_{HO}(\vec{r})$). The essential point now is that the transition from the known system to the unknown system ($U(\vec{r} - \vec{r}_i)$) can be performed by means of the time – dependent Schrödinger equation

$$i\hbar \frac{\partial \widetilde{\Psi}\left(\frac{\vec{r}}{R}, t\right)}{\partial t} = \left[ -\vec{\nabla}_{(\vec{r}/R)} + U_{HO}\left(\frac{\vec{r}}{R}\right) + \sigma(t) U\left(\frac{\vec{r} - \vec{r}_i}{R}\right) \right] \widetilde{\Psi}\left(\frac{\vec{r}}{R}, t\right) \qquad (5)$$

with $U_{HO}\left(\dfrac{\vec{r}}{R}\right) = \left(\dfrac{|\vec{r}|}{R}\right)^2$ (the usual harmonic oscillator in dimensionless units),

$U\left(\dfrac{\vec{r} - \vec{r}_i}{R}\right) = -\dfrac{2\left(\dfrac{R_y}{E_0}\right)^{1/2}}{\left|\dfrac{\vec{r} - \vec{r}_i}{R}\right|}$ (the Coulomb interaction between electron and donor in

dimensionless units) and $\sigma(t)$ has the property: $\sigma(t) = 0$, for $t \leq t_a$ and $\sigma(t) = 1$ for $t \geq t_b$. For $t_a \leq t \leq t_b$ the function $\sigma(t)$ may have any shape but should increase monotonically. A simple choice which we have used in our calculations in the present paper is $\sigma(t) = a(t - t_a)$ with $a = 1/(t_b - t_a)$.

Then the energy $\widetilde{E}$ (eq. 3) is obtained as follows

$$\widetilde{E} = \int \widetilde{\Psi}^*\left(\frac{\vec{r}}{R}\right) \left[ -\vec{\nabla}_{(\vec{r}/R)} + \left(\frac{|\vec{r}|}{R}\right)^2 - \frac{2\left(\dfrac{R_y}{E_0}\right)^{1/2}}{\left|\dfrac{\vec{r} - \vec{r}_i}{R}\right|} \right] \widetilde{\Psi}\left(\frac{\vec{r}}{R}\right) d\vec{r} \qquad (6)$$

where $\widetilde{\Psi}\left(\dfrac{\vec{r}}{R}\right) = \widetilde{\Psi}\left(\dfrac{\vec{r}}{R}, t\right)$ is the wavefunction of the system for $t \geq t_b$.

Finally, in order to calculate the binding energy $E_b$ of the hydrogenic impurity, the difference between the energy states without and with the impurity present for a particular level is estimated.

## 3. Results

For numerical calculation of the donor binding energies we concentrate on GaAs semiconductor and we use the material parameters from ref. [8,15].

For donors impurities in GaAs, the Rydberg constant $R_y$ is approximately 5.24meV, and the effective mass of the electron is $0.067\,m_e$. Using these parameters we get that $E_0 = \frac{108.52}{R^2} R_y$ and the coefficient of the electrostatic term is $0.192R$, where in both expressions the dot radius $R$ is expressed in nm (see Appendix).

*Two-dimensional systems*

In order to obtain the binding energy for the ground and few adjacent states in two-dimensional QDs with PMM we use as a reference system the usual harmonic oscillator in two–dimensions with eigenfunctions [16]

$$\Psi(r) = \lambda^{-1} u^{|m|/2} e^{-u/2} L_n^{|m|}(u) \tag{7}$$

where $u = r^2/2\lambda^2$, $\lambda^2 = \hbar/m^*\omega$, $L_n^{|m|}(u)$ is a Laguerre polynomial with $n = 0,1,2.$ and $m = 0, \pm 1, ...., \pm n$.

The binding energy of a (shallow) hydrogenic impurity in spherical GaAs QD with parabolic confinement is plotted as a function of the radius ($R$), for various positions of the impurity within the dot (Fig. 1a). We observe that the impurity binding energy decreases as the size of the QD increases. This is expected as an increase in the dot radius results in a spreading of the wave function which consequently causes lowering in the binding energy. Similar behavior is observed for excited states (Fig. 1b). As the size of the QD is increased the binding energy for different impurity positions ($r_i = 0$, $0.5R$) are found to converge to the same value (bulk limit). Moreover, as the curve which corresponds to $r_i = 0$ is always higher than the curve with $r_i = 0.5R$ (Fig. 1a) we conclude that the binding energy is maximum for an on-center impurity. Both figures have been plotted with $d^2_{cutoff} \equiv 0.1 < | \vec{r} - \vec{r}_i |^2 > - < | \vec{r} - \vec{r}_i |>^2$.

Similar results have been evaluated assuming fixed cut off distance (not shown). Results reported for the same system studied by perturbation method [8] and/or a variational approach [10] within the effective mass approximation show similar behavior.

In Figs. 2a and 2b we display the variation of the binding energies associated with the four lowest states in a QD as a function of the impurity position for different cutoff values. We see that for s-states ($m = 0$) e.g. (00) and (10), the maximum value of the binding energy is achieved for an on-center impurity. As we move away from the center of the dot the binding energy decreases smoothly. The observed variations can be attributed to the nature of zero order electronic density corresponding to the s-states. This is due to the fact that the zero-order electronic density, proportional to the wave function, has a maximum at the center of the dot. This became lower and wider when $n$ increases. The case of p ($m=1$) and d ($m=2$) states shows a nonmonotonic dependence on the position of the impurity. We observe a broad maximum shifted away from the center of the dot. The maximum of the d-state is located further away from the center of the dot as compared to that of the p-state. Similarly, the observed variations can be attributed to the nature of zero order electronic density corresponding to the p- and d-states.

All results mentioned till this point refer to $R_{dot}$=2.76nm. Similar results are observed for different sizes of the QDs as shown in Fig.3, assuming a fixed cut off distance, $d^2_{cutoff}$=$10^{-5}R^2$. More specifically, in Fig. 3a, which corresponds to smaller dot radius ($R_{dot}$=1nm) the values of energy for all the states except the d state increase due to larger confinement by a factor close to two compared to results in Fig. 2a. Furthermore, for weak confinement $R_{dot}$=15nm, the energy values for all states get suppressed but they are less sensitive to the position of the impurity.

*Three - dimensional systems.*

In order to obtain the binding energy for the ground and few adjacent states in three - dimensional QDs with PMM we use as a reference system the usual harmonic oscillator in three – dimensions with eigenfunctions [17]

$$\Psi_{nlm}(r,\theta,\phi) = r^l e^{-(\lambda/2)r^2} {}_1F_1\left(-n, l+3/2; \quad \lambda\, r^2\right) Y_{lm}(\theta,\phi) \qquad (8)$$

The binding energy of a shallow hydrogenic impurity in spherical QD with parabolic confinement is plotted as a function of the impurity position within the dot in Fig.4 for $d_{cutoff}^2 = 10^{-5}R^2$ (the case with $d_{cutoff}^2 = 10^{-5}R^2$ (fixed), is very close to the $0.1\left(<|\vec{r}-\vec{r}_i|^2> - <|\vec{r}-\vec{r}_i|>^2\right)$ case e.g. the difference in energy is less than 0.2 R$_y$). In Figs. 4, 5 we report results for a QD with radius $R$=2.76nm [8].

We see in Fig. 4, that for s-states ($l = 0$) e.g. (000) and (100), the maximum value of the binding energy is achieved for an on-center impurity. This is due to the fact that the zero-order electronic density, proportional to the wave function, has a maximum at the center of the dot. As we move away from the center of the dot the binding energy decreases smoothly. This became lower and narrower when $n$ increases. Moreover we study $n = 0$ states with non-zero $l$ values. The case of p ($l$=1) and d ($l$=2) states shows a nonmonotonic dependence on the position of the impurity. In addition we observe a broad maximum shifted away from the center of the dot. The maximum of the d-state is located further away from the center of the dot as compared to that of the p-state. The observed variations are purely attributed to the nature of zero order electronic density corresponding to the p- and d-states, which show maxima at different positions. From the wavefunctions of the three-dimensional harmonic oscillator (eq.8), we expect that the ratio of the positions of the distinct maxima should be $\sqrt{2}$. Actually this is what is observed in Fig. 4, where the position of the maximum for p-state is around $0.98R$ and the position of the maximum for the d-state is around $1.4R$. This indicates that the wavefunction of the Hamiltonian with the Coulomb interaction (eq. 1) is not very different from the wavefunction of the 3D harmonic oscillator.

Furthermore as is shown in Fig. 5 for the p-states which are triply degenerate, the Coulomb interaction lifts the degeneracy due to which the donor state corresponding to p-states splits into two levels. These results are in an excellent agreement with the corresponding results of ref. [8].

## 4. Conclusion

We have presented a calculation for the donor binding energies associated with the ground state and a few excited states in two and three dimensional GaAs QDs with parabolic confinement. The calculation has been performed by means of recently developed numerical method (PMM). The computed results show that the impurity binding energies increase with the decrease in dot size and that for s − states the binding energy has a maximum for an on − center impurity, while for p and d states the maximum occurs for an impurity located off the dot center. Furthermore, for the three dimensional case the impurity binding energies of the p- levels are found to split as the location of the donor is varied within the QD. The results we have obtained are in an excellent agreement with the corresponding results of perturbation and variational method.

Our method is rather general as it can be applied to any confinement limits (from weak (Fig. 3b) to strong confinement (Fig. (3a)) in contrast to the usual perturbation method which is working only in the strong confinement limit.

### Acknowledgements

The authors would like to thank Prof. W. Schommers and Dr. M. Rieth for fruitful discussions. A.F.T. and S. B. would like to thank also the Research Committee of University of Patras, Greece, for financial support under the project Karatheodoris.

### Appendix

Lets $R$ be the radius of the spherical QD and $E_0$ reference energy both not known at the moment. We divide both sides of eq.2 by $E_0$ and introduce the dot size in the right hand side of equation by multiplying the two first terms by $R^2/R^2$ and the last term by $R/R$. eq.2 gives:

$$\left[ -\frac{\hbar^2}{2mR^2 E_0} \left( \frac{\partial^2}{\partial \left(\frac{x}{R}\right)^2} + \frac{\partial^2}{\partial \left(\frac{y}{R}\right)^2} + \frac{\partial^2}{\partial \left(\frac{z}{R}\right)^2} \right) + \frac{m\omega^2 R^2}{2E_0} \left(\frac{r}{R}\right)^2 - \frac{e^2}{\varepsilon R E_0 \left| \frac{\vec{r}}{R} - \frac{\vec{r}_i}{R} \right|} \right] \Psi(\vec{r}) = \frac{E}{E_0} \Psi(\vec{r})$$

$$(A1)$$

In order to simplify things we look for values of $E_0$ and $R$ such that the coefficients of the kinetic and confinement term are unity (i.e. $\frac{\hbar^2}{2mR^2 E_0} = 1$ and $\frac{m\omega^2 R^2}{2E_0} = 1$). The solution of these two equations gives $E_0 = \frac{\hbar\omega}{2}$ and $R = \left(\frac{\hbar}{m\omega}\right)^{1/2}$.

Hence the coefficient of the electrostatic term can be simplified to:

$$\frac{e^2}{\varepsilon R E_0} = \frac{e^2}{\varepsilon \sqrt{\frac{\hbar}{m\omega}} \frac{\hbar\omega}{2}} = 2\sqrt{\frac{me^4}{2\hbar^2 \varepsilon^2} \frac{1}{(\hbar\omega/2)}} = 2\sqrt{\frac{R_y}{E_0}} ,$$

where $R_y$ is the Rydbergs's constant.

Therefore eq. (A1) can be written in the form of eq. 3.

**Figure Captions**

**Fig. 1a.** Plots of the binding energy for the 00 state of a hydrogenic impurity in a two dimensional parabolic GaAs QD as a function of the radius ($R$), for various positions of the impurity within the dot: $r_i$=0 (squares) and $r_i$=0.5$R$ (circles).

**Fig. 1b.** Plots of the binding energy of a hydrogenic impurity in a two dimensional parabolic GaAs QD as a function of the radius ($R$), for two states: 00 and 01 at $r_i$=0.

**Fig. 2.** Plots of the binding energy of a hydrogenic impurity in a two dimensional parabolic GaAs QD as a function of the impurity location for QD radius R = 2.76 nm using various cut off distances $d_{cutoff}^2$ . **a.** $d_{cutoff}^2$ =10$^{-5}R^2$. **b.** $d_{cutoff}^2 \equiv 0.1 < |\vec{r} - \vec{r}_i|^2 > - < |\vec{r} - \vec{r}_i| >^2$ .

**Fig. 3.** Plot of the binding energy of a hydrogenic impurity in a two dimensional parabolic GaAs QD as a function of the impurity location, for cut off distance $d_{cutoff}^2$ = 10$^{-5}$ $R^2$ and for QD radius **a.** $R$ = 1 nm and **b.** $R$ = 15nm.

**Fig. 4.** The binding energy of the four lowest donor states (000, 100, 010 and 020) as a function of the impurity position within a three dimensional parabolic GaAs QD for cut off distance $d_{cutoff}^2$ = 10$^{-5}$ $R^2$. The QD radius is R = 2.76 nm.

**Fig. 5**. The splitting of the donor binding energies of the p- states (010 and 011) as a function of the impurity position within a three dimensional parabolic GaAs QD for cut off distance $d_{cutoff}^2$ = = 10$^{-5}$ $R^2$. The QD radius is R = 2.76 nm.

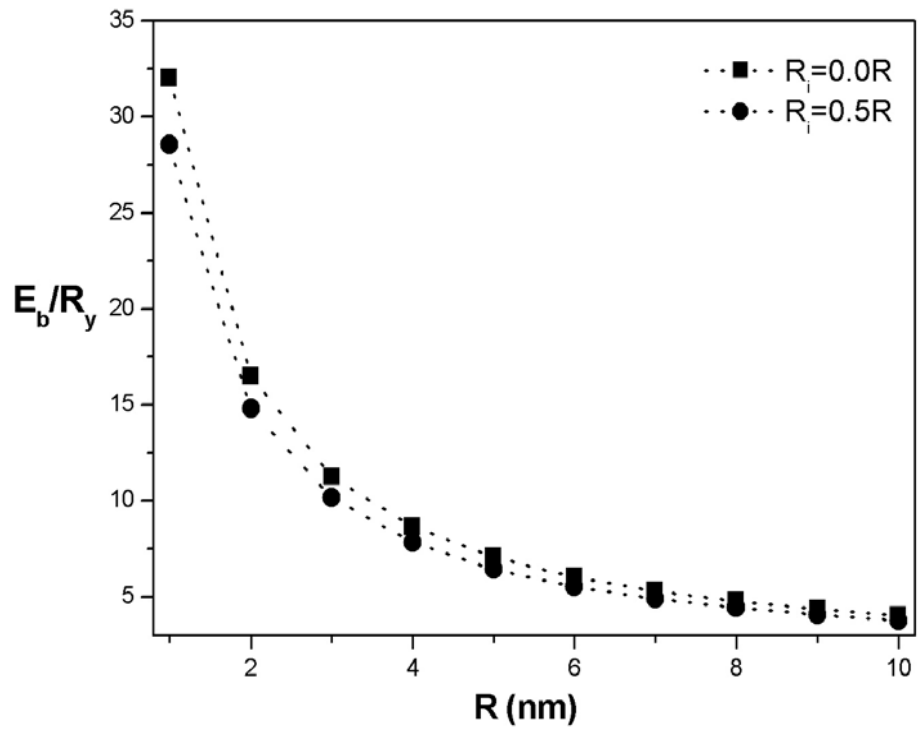

**Fig. 1a.**

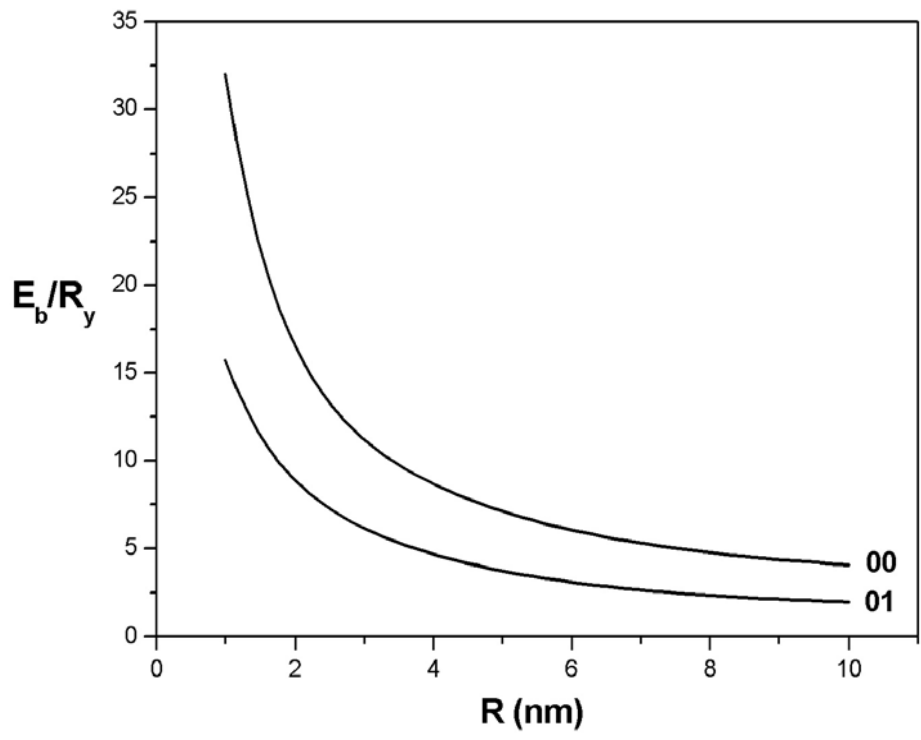

**Fig. 1b.**

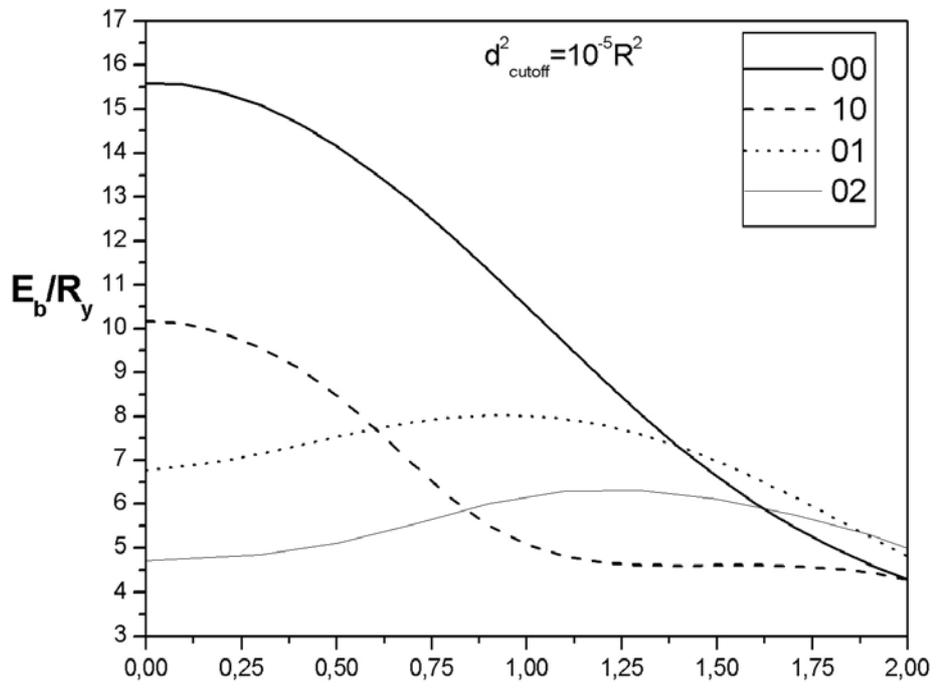

**Fig. 2a.**

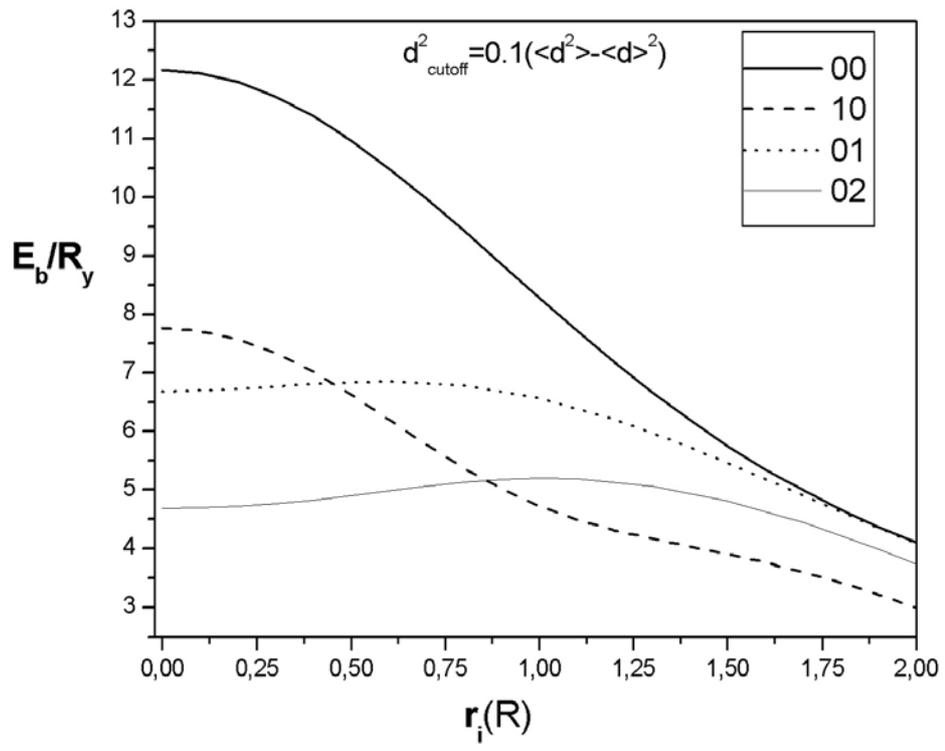

**Fig. 2b.**

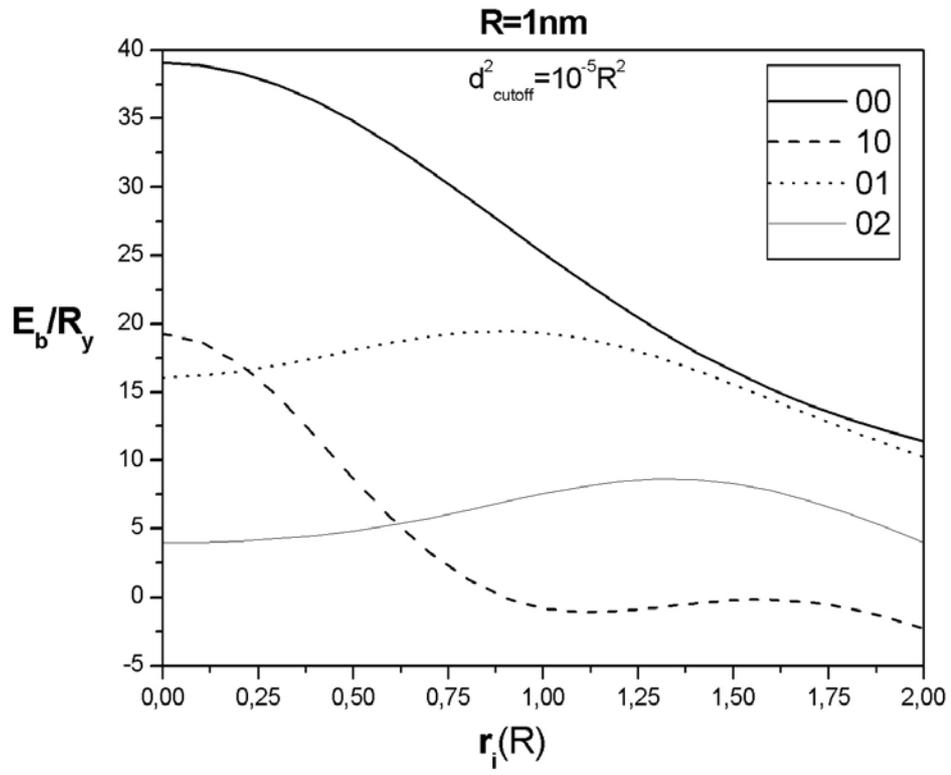

**Fig. 3a.**

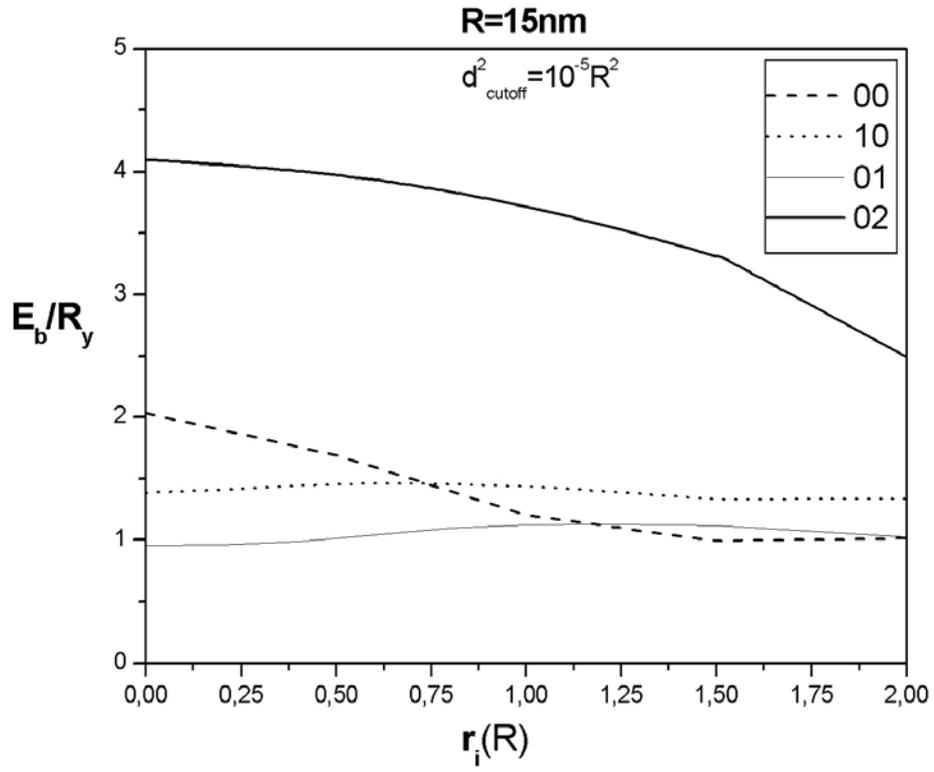

**Fig. 3b.**

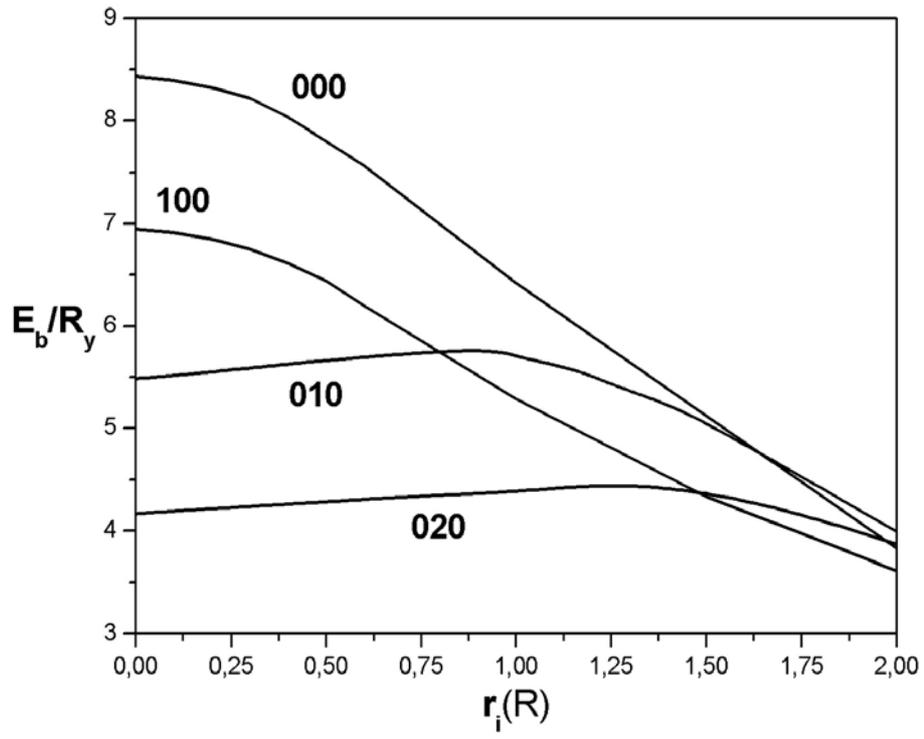

**Fig. 4.**

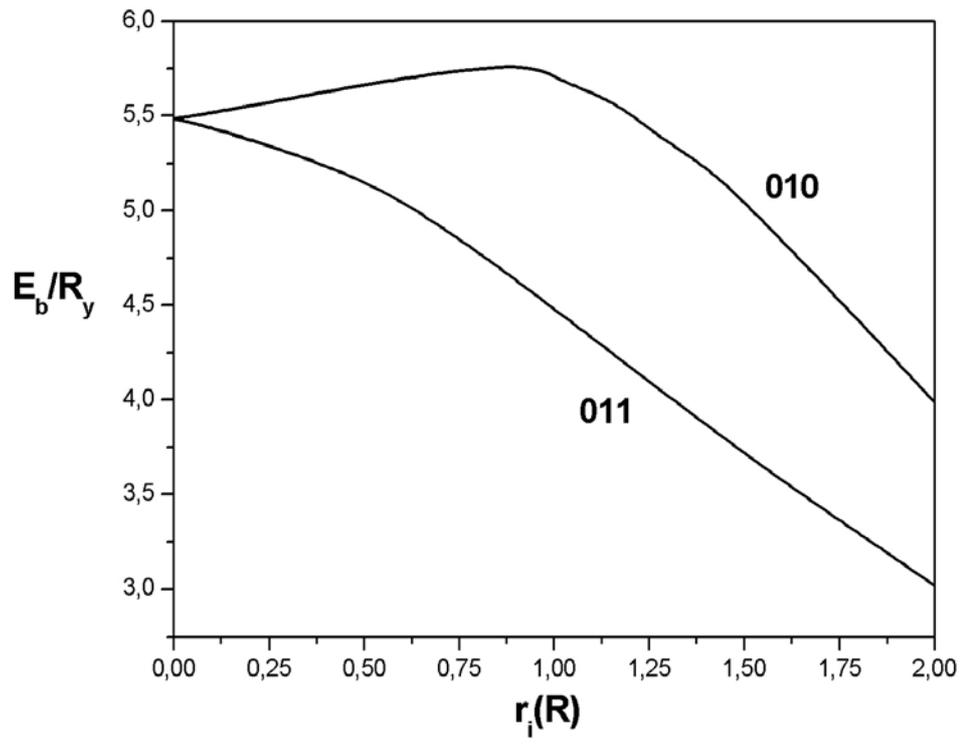

**Fig. 5.**